# NSV 13983: A NEW DWARF NOVA IN THE PERIOD GAP⋆
## *(Research Note)*

C. Contreras and C. Tappert

Departamento de Astronomía y Astrofísica, Pontificia Universidad Católica, Casilla 306, Santiago 22, Chile

**ABSTRACT**

*Aims.* NSV 13983 is catalogued as a dwarf nova based on a reported outburst from 2005. The system has not yet been studied spectroscopically. We attempt to confirm its nature as a dwarf nova and determine its orbital period.
*Methods.* We derive the orbital period by using time-resolved spectroscopic data to measure radial velocities.
*Results.* The average spectrum shows evidence that the system is a dwarf nova in quiescence. The radial velocity curves derived from measurements of the spectral lines H$\alpha$ and H$\beta$, show a clear modulation with a period of 2.76 h. This places NSV 13983 below the upper edge of the gap in the period distribution of cataclysmic variables, implying that it is the 14th dwarf nova in the gap.

## 1. Introduction

Cataclysmic Variable stars (CVs) are close interacting binary systems that comprise a white-dwarf primary that accretes material from a main-sequence secondary. Dwarf novae (DNe) represent a subclass of CVs for which the process of accretion occurs via an accretion disc, which undergoes episodes of brightness variations observed as outbursts. According to the morphology of long-term variations, DNe can be classified as SU UMa, U Gem, or Z Cam systems. For a comprehensive review of CVs see Warner (1995).

The evolutionary status of CVs is believed to be strongly related to the orbital period, which therefore represents one of the most important system parameters. The period distribution of CVs consequently qualifies as an observational testbed for theoretical models of CV evolution. One of its characteristic features is the apparent lack of systems with periods ranging between 2-3 h. This so-called period gap represents a division between high-mass-transfer, long-period systems, that are mainly driven by magnetic braking, and CVs of short orbital periods and low-mass-transfer rates (Kraft et al. 1962; Verbunt & Zwaan 1981).

NSV 13983 was first observed to be a potential variable star at Bamberg Observatory by Friedrich & Schoffel (1971), and measured to have a photographic magnitude of 11. The system is classified as a dwarf nova in the Downes et al. (2005) catalogue based on an outburst reported during an AAVSO discussion in April 2005. We present the first spectroscopic study of NSV 13983, and derive its orbital period based on the radial-velocity variation in the system emission lines.

## 2. Observations and data reduction

We obtained time-resolved spectroscopic data on July 4, 2006 at CTIO using the R-C spectrograph mounted on the Blanco 4 m telescope. The KPGL3 grating was used in combination with a 1.0” slit yielding a wavelength range of ~3590-7290 Å and a spectral resolution of 4.2 Å.

Twenty-four exposures were taken covering 3.08 h of observations. Individual exposure times were 240 s and 300 s. In addition, HeNeAr lamp spectra were taken every 5 object exposures for wavelength calibration purposes. No observations were acquired to complete a flux calibration of the data. Bias correction, flatfielding, and wavelength calibration of spectra were performed using the standard IRAF [1] routines.

## 3. Results

### 3.1. Average spectrum

The average spectrum of NSV 13983 is presented in Fig. 1 and the properties of its prominent features are given in Table 1. The spectrum is dominated by strong emission lines of the Balmer series, clearly visible from H$\alpha$ through to H$\epsilon$, and also by He I lines. We can furthermore identify Ca II (H and K emission) and Fe II $\lambda$5169 lines. We do not observe highly-ionized lines like He II nor any lines that could be associated with the secondary star. The strength of the emission lines and the absence of HeII, imply that the system is a non-magnetic low-mass-transfer system, i.e. a dwarf nova. The absence of clear signatures of the secondary star reflects the faintness of this component and indicates a short orbital period.

### 3.2. Radial velocities

To determine the orbital period for this system, the radial velocity variations of the H$\alpha$ and H$\beta$ lines were determined by fitting a single Gaussian profile to each emission line using the “k” routine in the splot package of IRAF. The measured radial velocities had to be corrected for two important effects: first, for the motion with respect to the local standard of rest, which was calculated using the rvcorrect routine in the astutil package of IRAF, and second, for the instrumental flexure between two calibration exposures by measuring the position of the night sky emission line at 6300.304 Å and subtracting the variation with respect to its rest wavelength.

⋆ Based on observations collected at the NOAO Cerro Tololo Interamerican Observatory, Chile

[1] IRAF is distributed by the national Optical Astronomy Observatories

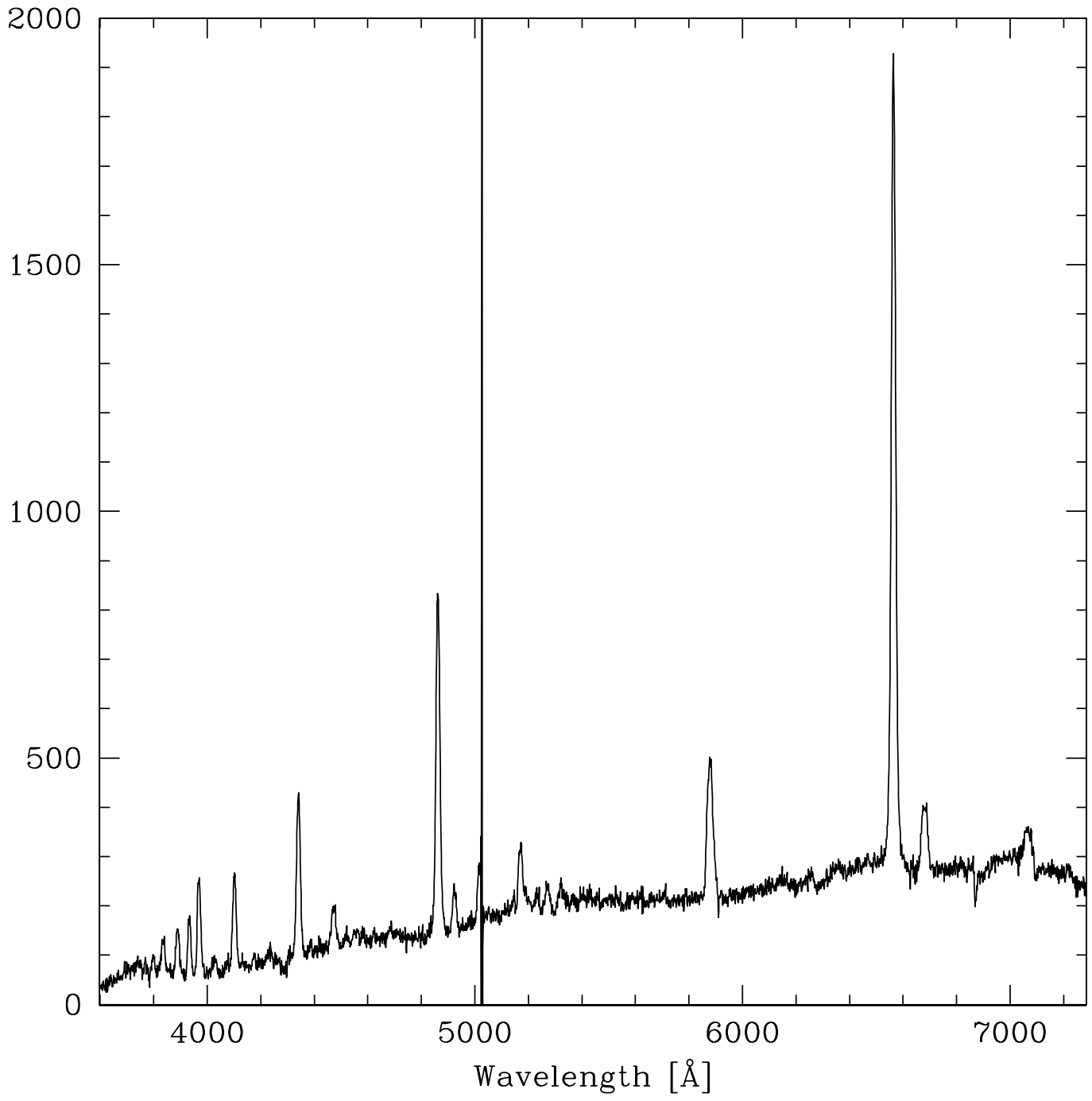

**Fig. 1.** Average spectrum of NSV 13983. The large artifact at ∼ 5016 Å could not be corrected due to its proximity to a He I line.

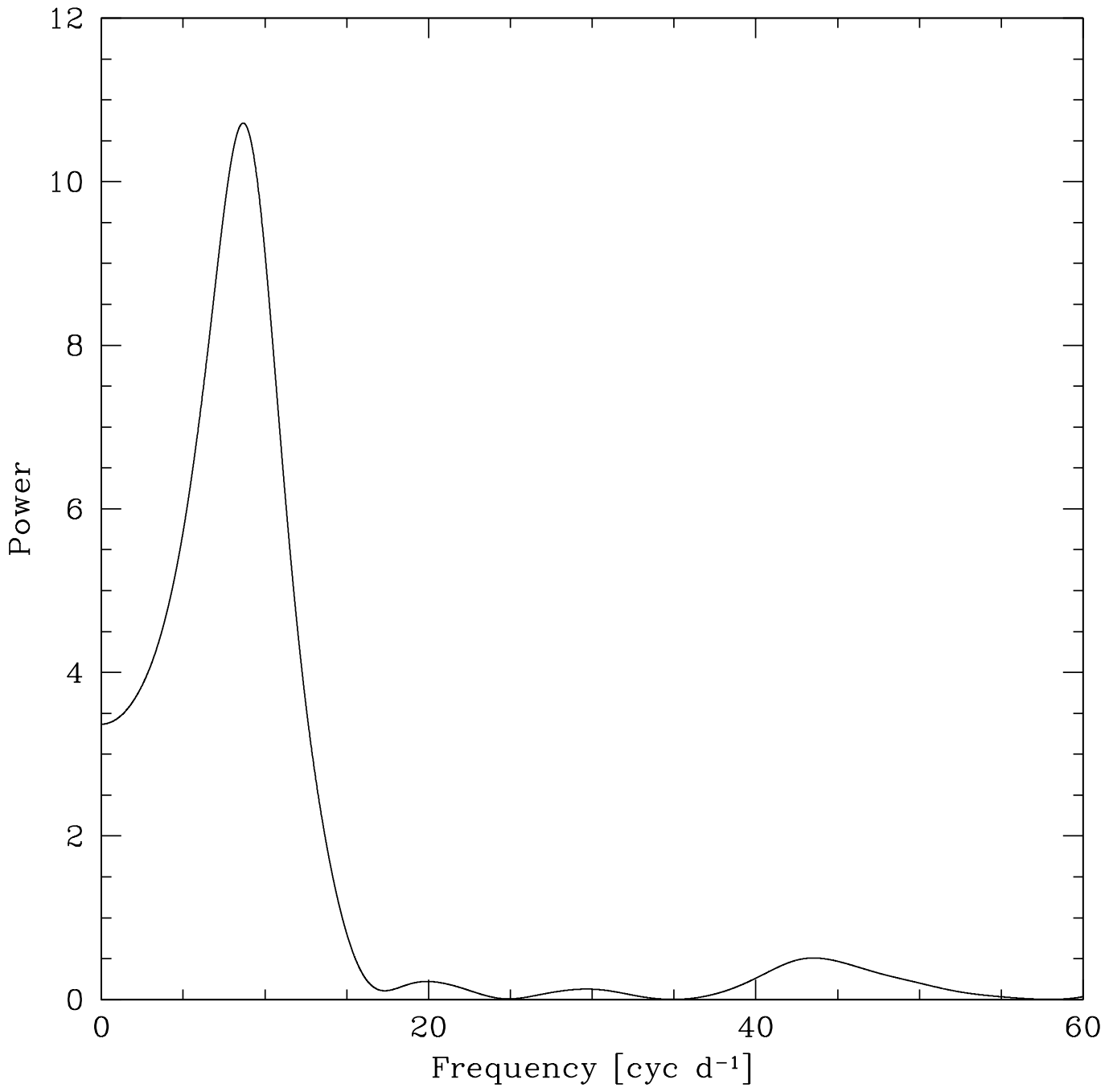

**Fig. 2.** Scargle periodgram for H$\alpha$. The peak corresponds to $P$ = 2.76 h.

**Table 1.** Properties of the most prominent features in the average spectrum of NSV 13983. Column 1 represents the observed wavelength determined by a Gaussian fit, column 2 provides the equivalent width, columns 3 and 4 the line identification and the rest wavelength, respectively. A colon marks uncertain values.

| (1) | (2) | (3) | (4) | (5) |
|---|---|---|---|---|
| $\lambda$ [Å] | $W_\lambda$ [Å] | Identification | $\lambda_0$ [Å] | remarks |
| 3798 | 7: | H10 | 3798 | [1] |
| 3835 | 18: | H9 | 3835 | [1] |
| 3889 | 23: | H8 + He I | 3889 | [1] |
| 3933 | 26: | Ca II K | 3934 | [1] |
| 3969 | 51: | H$\epsilon$ + Ca II H | 3970 | [1] |
| 4027 | 12 | He I | 4026 | |
| 4102 | 41 | H$\delta$ | 4102 | |
| 4341 | 60 | H$\gamma$ | 4341 | |
| 4473 | 14 | He I | 4471 | |
| 4862 | 95 | H$\beta$ | 4861 | |
| 4924 | 10 | He I | 4922 | |
| 5017 | 13: | He I | 5016 | [2] |
| 5171 | 12 | Fe II | 5169 | |
| 5325 | 8 | Fe II | 5317 | [3] |
| 5879 | 36 | He I | 5876 | |
| 6564 | 117 | H$\alpha$ | 6563 | |
| 6680 | 14 | He I | 6678 | |
| 7065 | 9 | He I | 7065 | |

[1] Lines are blended.
[2] Line is distorted due to a CCD artifact
[3] ID uncertain

Due to its superior signal-to-noise ratio, only radial velocities determined using the emission line of H$\alpha$ were analyzed to search for periodicity. This analysis was performed using the Scargle algorithm (Scargle 1982) implemented in MIDAS. Due to the limited orbital coverage of only one night of observations, the periodogram yields one broad peak that is centered on $f$ = 8.68 cyc d$^{-1}$, corresponding to $P$ = 2.76 h (Fig. 2).

The radial velocities of both emission lines were folded by this period and fitted by the sine function

$$v(\varphi) = \gamma - K_1 \sin(2\pi\varphi), \tag{1}$$

where $v$ is the measured radial velocity and $\varphi$ is the phase with respect to the period $P$. The fit parameters are the offset of the modulation with respect to the observer $\gamma$ and the semi-amplitude $K_1$. In this case, we take $\varphi_0$ to represent the first observation (HJD = 2453921.8069). A Monte-Carlo method was used to estimate the uncertainties in the parameters of the fit. This process started by fitting the data with a Fourier series and determining the parameters of this fit. Next, we varied the radial velocities by an amount randomly selected from a range centered on the actual measurement of width $2\sigma$, fitted the varied data and determine a fresh set of parameters. The process was repeated for ten thousand iterations. Finally, we estimated the uncertainties in the original parameters to be the standard deviation of parameters determined from Monte Carlo simulations. In Table 2, we present the results of these fits for both H$\alpha$ and H$\beta$.

The uncertainty in the period, $\sigma_P$, was estimated by using Eq. 4 of Larsson (1996),

$$\sigma_P^2 = \frac{6\sigma_{tot}^2}{\pi^2 N A^2 T^2} P^4 \tag{2}$$

where $\sigma_{tot}$ corresponds to the standard deviation of the radial velocity data, $N$ is the number of data points, $A$ represents the amplitude of the oscillation, and $T$ is the total time covered by the observations. This yields $\sigma_P$ = 0.27 h.

The corresponding radial velocities and their fits are presented in Fig. 3. The H$\beta$ plot shows a larger scatter of the data points compared to that of H$\alpha$, but still yields an acceptable fit with respect to $P$. We therefore identify $P$ = 2.76(27) h with the orbital period of NSV 13983. Nevertheless, because of the limited amount of data presented here, more observations would be desirable to be able to determine a more accurate constraint on this value.

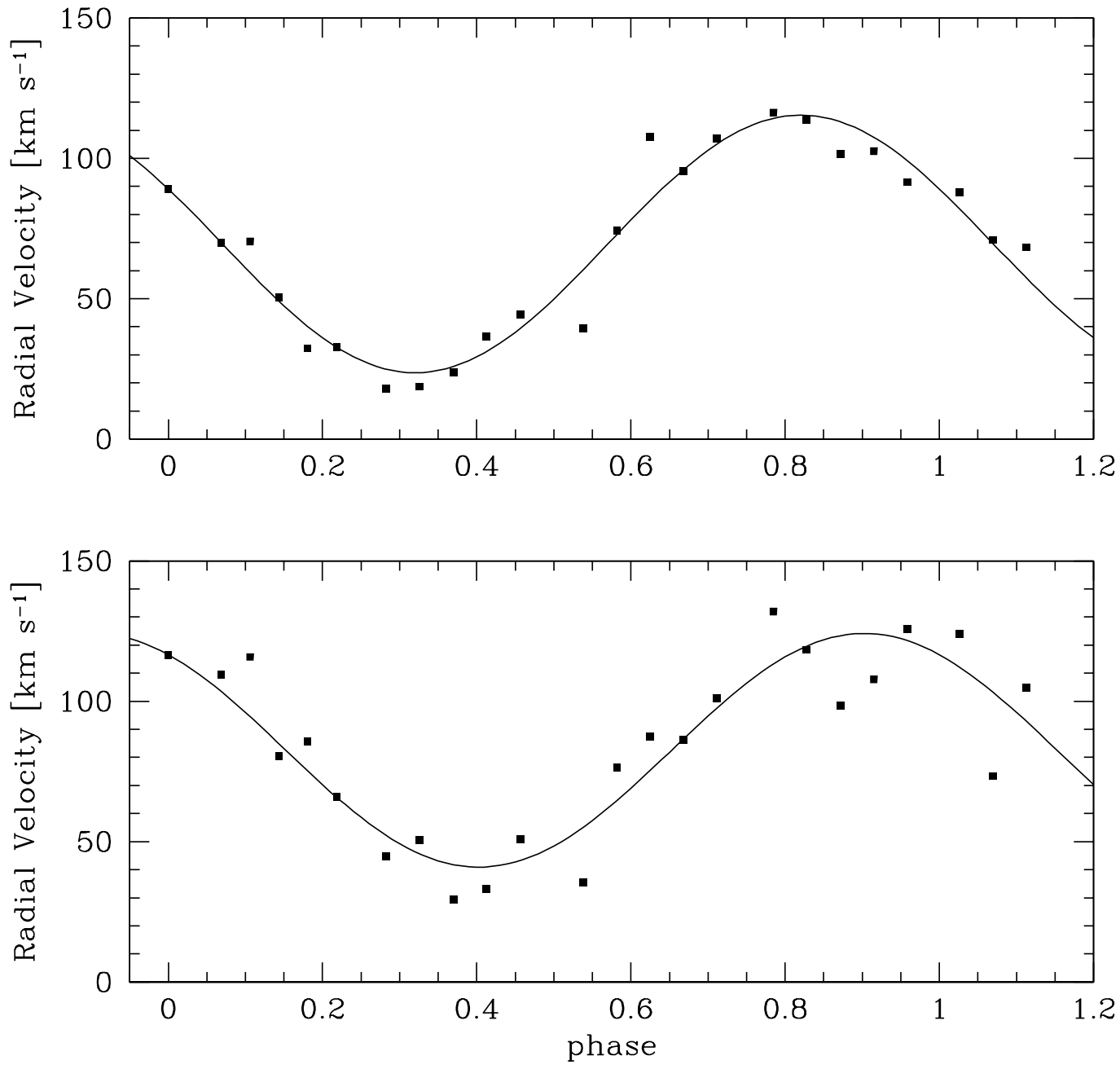

**Fig. 3.** Radial velocities of H$\alpha$ (*top*) and H$\beta$ (*bottom*) versus orbital phase. The data are represented in the order of their observation, i.e. phase 0 is set to be the first observed data point, and the last observation corresponds to phase 1.11. The solid curve represents the best fit with respect to $P$ = 2.76 h.

**Table 2.** Calculated fit parameters

| Line | $\gamma$ [km s$^{-1}$] | $K_1$ [km s$^{-1}$] | $\sigma$ [km s$^{-1}$] |
|---|---|---|---|
| H$\alpha$ | 69.52 ± 1.02 | 45.89 ± 1.70 | 8.44 |
| H$\beta$ | 82.52 ± 1.62 | 41.62 ± 2.32 | 13.30 |

## 4. Conclusion

We speculate that NSV 13983 has a low orbital inclination, due to the low semi-amplitude $K_1$ = 46 km s$^{-1}$ of its radial velocity. In contrast, Szkody et al. (2004) measured $K_1$ = 110 km s$^{-1}$ for the radial velocity of H$\alpha$ in SDSS J170213.26+322954.1, also a dwarf nova in the gap. This system has a high orbital inclination of $i$ = 82.4° (Littlefair et al. 2006).

The derived spectroscopic period places NSV 13983 in the upper third of the period gap (Fig. 4), where we have assumed the limits on the period gap to be $P$ = 3.18 h (upper edge) and $P$ = 2.15 h (lower edge) given by Knigge (2006).

Schmidtobreick & Tappert (2006) listed 11 dwarf novae of periods that reside in the gap. One more system, V1006 Cyg, is added in version 7.9 of the Ritter & Kolb (2003) catalogue. Dillon et al. (2008) found a period of 2.38 h for the possible dwarf nova SDSS J124426.26+613514.6. This makes NSV 13983 the 14th dwarf nova to be found in the period gap and the one with the fourth longest orbital period in the gap.

*Acknowledgements.* We would like to thank the referee for helpful comments

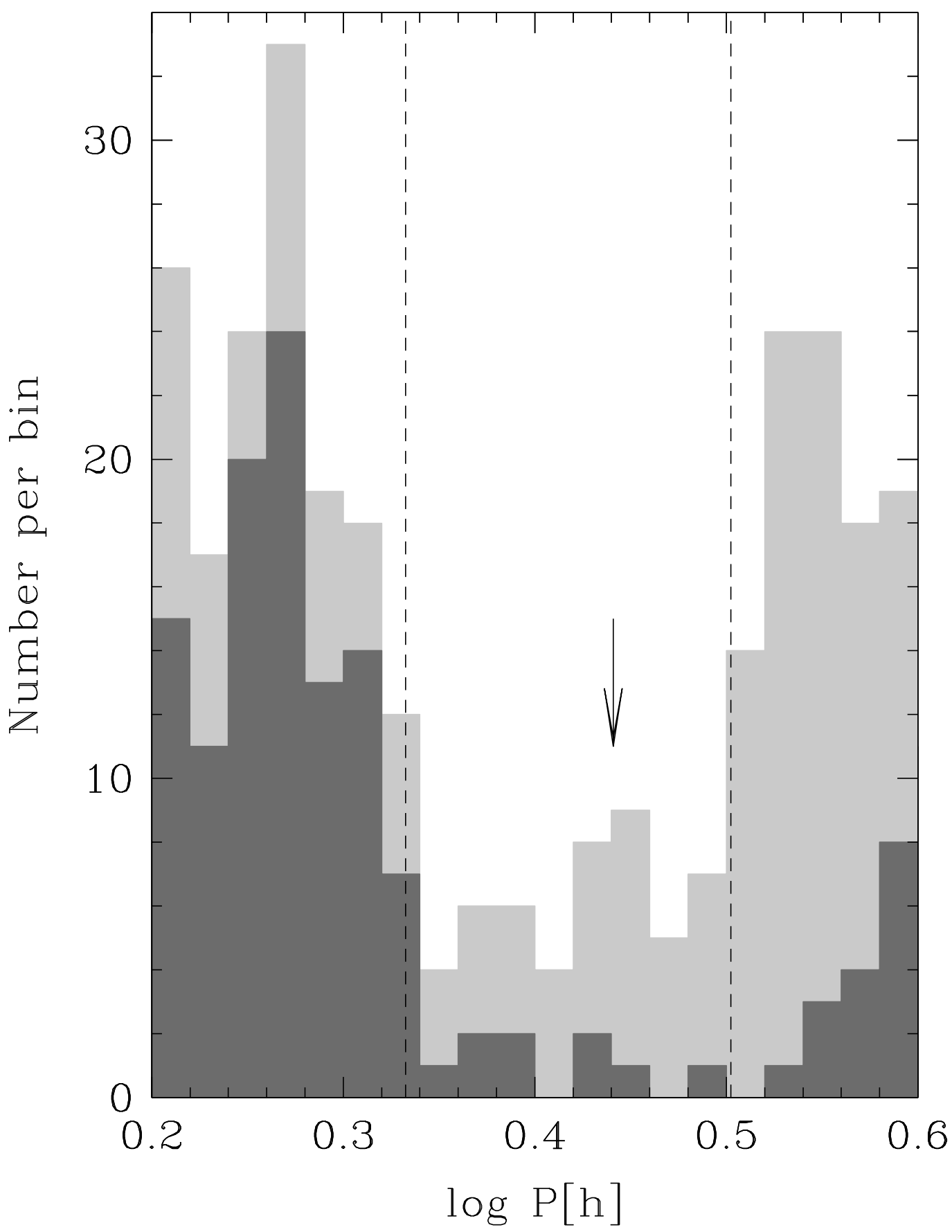

**Fig. 4.** Close-up of the period distribution of CVs about the period gap, based on version 7.9 of the Ritter & Kolb (2003) catalogue, excluding objects of uncertain period (i.e., those marked with a colon). Two distributions are shown: all CV subtypes (light grey) and dwarf novae only (dark grey). The dashed lines mark the limits of the period gap defined by Knigge (2006). The bin size is 0.02 units. The arrow marks the position of NSV 13983, which is not included in the histograms.